\documentclass[12pt]{article}
\usepackage{epsfig}
\usepackage{amssymb}
\usepackage{amsmath}
\usepackage{latexsym}

\newcommand{\ttt}{t\bar{t}}
\newcommand{\alphas}{\alpha_{s}}

\def\beq{\begin{equation}}
\def\eeq#1{\label{#1}\end{equation}}
\def\eeqn{\end{equation}}

\def\beqa{\begin{eqnarray}}
\def\eeqa#1{\label{#1}\end{eqnarray}}
\def\eeqan{\end{eqnarray}}

\let\bar=\overbar

\def\Dslash{\not{\hbox{\kern-4pt $D$}}}
\def\dslash{\not{\hbox{\kern-2pt $\del$}}}

\def\alphas{\alpha_s}
\def\msb{{\bar{\ssstyle M \kern -1pt S}}}

\def\Title#1{\begin{center} {\Large {\bf #1} } \end{center}}

\begin{document}

\Title{{\footnotesize\vspace*{-3.5cm}\hspace*{12cm} MPP-2012-157}\vspace*{1cm}\\The electroweak contribution to top quark pair production:
cross sections and asymmetries}
\bigskip
\footnotesize Proceedings of CKM 2012, the 7th International Workshop on the CKM Unitarity Triangle, University of Cincinnati, USA, 28 September - 2 October 2012 \normalsize\bigskip
\begin{raggedright}  

{\it Davide Pagani\index{Pagani, D.}\\
Max-Planck-Institut f\"{u}r Physik\\
F\"{o}hringer Ring 6\\
80805 M\"{u}nchen, GERMANY\\}
\bigskip
\end{raggedright}
\section{Introduction}
\index{Electroweak corrections}
\index{Top quark pair production}
The production of top quark pairs at hadron colliders is, first of all, a QCD process. 
At LO $(\mathcal{O}(\alpha_{s}^{2}))$, the process originates via strong interactions from two different partonic subprocesses: $q\bar{q}\rightarrow\ttt$ and $gg\rightarrow\ttt$. The NLO QCD corrections $(\mathcal{O}(\alpha_{s}^{3}))$ have to be taken into account to obtain a realistic phenomenological description and reduce the errors due to the variation of the renormalization and factorization scale. These errors can be further reduced including NNLL  corrections \cite{Ahrens:2010zv,Kidonakis:2010dk,Beneke:2011mq,Cacciari:2011hy,Kidonakis:2012rm}. Also complete NNLO calculations for the $q\bar{q}\rightarrow\ttt(+X)$ \cite{Baernreuther:2012ws} and $qg\rightarrow\ttt(+X)$ \cite{Czakon:2012pz} channels, NLO factorizable corrections including the decay in narrow-width-approximation \cite{Melnikov:2009dn,Campbell:2012uf} and the complete NLO including leptonic decays \cite{Bevilacqua:2010qb,Denner:2012yc} are known.

Besides pure QCD effects, also electroweak interactions induce corrections to $\ttt$ production \cite{Beenakker:1993yr,Kuhn:2006vh,Bernreuther:2006vg,Hollik:2007sw,Bernreuther:2010ny}. The size of these corrections and their phenomenological impact strongly depend on the particular physical observable under consideration.
The asymmetric part of the cross section, arising in QCD at $\mathcal{O}(\alpha_{s}^{3})$, yields a charge asymmetry \cite{Kuhn:1998kw,Ahrens:2011uf,Brodsky:2012ik}. The relative  electroweak corrections \cite{Hollik:2011ps,Kuhn:2011ri,
Manohar:2012rs,Bernreuther:2012sx} to the QCD charge asymmetry prediction are much larger than in the case of the total cross section. However, also differential cross sections involving a very hard top exhibit large corrections. 
In this talk we discuss these different aspects of electroweak corrections to top quark production. We focus the attention on the contributions from $\mathcal{O}(\alpha_{s}^{2}\alpha)$ corrections, that can be divided into the QED part (only QED and QCD interactions) and the weak part (QCD and pure weak interactions). In some cases we include also $\mathcal{O}(\alpha^{2})$ corrections.
\section{Cross section}
\index{Top quark pair cross section}
The weak $\mathcal{O}(\alpha_{s}^{2}\alpha)$ corrections to the total cross section have been calculated for the first time in \cite{Beenakker:1993yr} and later at differential level in \cite{Kuhn:2006vh} and including polarizations and spin correlations in \cite{Bernreuther:2006vg}. At this order, the $gg\rightarrow\ttt(+X)$ process contributes only with electroweak loop corrections. The $q\bar{q}\rightarrow\ttt(+X)$ process contributes with loop corrections and also with the real gluon radiation via the interference of  $q\bar{q}\rightarrow Z \rightarrow\ttt g$ and $q\bar{q}\rightarrow g \rightarrow\ttt g$ diagrams. 

With the Higgs mass of order 126 GeV, weak corrections amount to $-2\%$ of the LO cross section, for the LHC at 14 TeV, and few permille  for the Tevatron.
These effects are small, especially if compared with the theoretical errors from PDFs and from the scale variation. In \cite{Ahrens:2010zv} these errors are estimated in different approximations and the NLO+NNLL accuracy presents the smallest values. At the LHC 14 TeV the error from PDFs is $\pm 3\%$ and the error from the variation of the scales\footnote{See \cite{Ahrens:2010zv} for more details on the procedure used to obtain the errors.} is $\pm 5\%$. At the Tevatron, they amount to $\pm 3\%$ and $\pm 4\%$, respectively.

The total cross section at the LHC gets the dominant contribution from the $gg$ initial state, whereas at the Tevatron the $q\bar{q}$ initial state contribution is larger than in the $gg$ case.
Conversely, the tail of the distributions involving a very hard top is dominated, for both the collider, by the $q\bar{q}$ initial state. In this regime, especially for the dominant $q\bar{q}\rightarrow \ttt(+X)$ process,  Sudakov logarithms enhance the size of relative electroweak corrections. 
For example, in the $p_{T}$ distribution at the LHC 14 TeV, they reach $-20\%$ for $p_{T}$ around 2~TeV. This effect is larger than the uncertainties discussed before. However, the absolute value of the differential cross section in the tail is six orders of magnitude smaller than in the peak region, so a very high luminosity is necessary to check these corrections with experimental data.
In \cite{Kuhn:2006vh} a similar quantity is discussed: the relative weak corrections to the LO cross section with a cut $p_{T}>p^{cut}_{T}$ applied. For $p^{cut}_{T}<2~\text{TeV}$ these corrections are larger than the statistical uncertainty estimated for an integrated luminosity of 200 $\text{fb}^{-1}$. It seems that it will be possible to test experimentally the effects from weak radiative corrections. The same arguments apply to the LHC at 7 and 8 TeV and the Tevatron, but the luminosity accumulated in these runs is not enough for a comparison with data \cite{Aad:2012hg}.
\\
\\
The remaining part of the $\mathcal{O}(\alpha_{s}^{2}\alpha)$ electroweak contribution, the QED corrections \cite{Hollik:2007sw}, involves only diagrams with QED and QCD interactions. 
The $gg\rightarrow \ttt(+X)$ process contributes, as in the weak case, with only loop corrections, the $q\bar{q}\rightarrow \ttt(+X)$ process with loop corrections, the real gluon radiation and also the real photon radiation.
In $q\bar{q}\rightarrow \ttt(+X)$, the boxes and the interference of the initial state and final state radiation do not contribute to the inclusive total cross section. Oppositely, the non vanishing terms do not contribute to the asymmetric cross section. The QED $\mathcal{O}(\alpha_{s}^{2}\alpha)$ of $q\bar{q}\rightarrow \ttt(+X)$ can be divided, on a diagrammatic base, into corrections to the total cross section and corrections to the charge asymmetry. Moreover, also the $\gamma g\rightarrow \ttt$ process is present at this order and, at the LHC 14 TeV, constitutes the dominant contribution to the QED corrections. Unfortunately, MRST2004QED \cite{Martin:2004dh} is at present the only PDF set that provides a photon distribution and allows in general to perform consistent calculations of electroweak corrections at hadron colliders.

The QED corrections to the total cross section, as in the weak case, are small: -2\% for the Tevatron and 1\% for the LHC at 14 TeV. Corrections to the distributions, again,  are enhanced in the regions involving a very hard top, but here they are reduced by the positive contribution of $\gamma g\rightarrow \ttt$. As illustrative comparison, in the $p_{T}$ distribution they are equal to -4\% for $p_{T}=2~\text{TeV}$. 

In \cite{Bernreuther:2010ny} NLO QCD results are combined with both QED and weak $\mathcal{O}(\alpha_{s}^{2}\alpha)$ corrections  and  also with contributions from $\mathcal{O}(\alpha_{s}\alpha)$, $\mathcal{O}(\alpha^{2})$ and $\mathcal{O}(\alpha_{s}\alpha^{2})$.
\section{Charge Asymmetries}
\index{Top quark charge asymmetry}
Both at the Tevatron and at the LHC, different definitions \cite{Aaltonen:2011kc,Abazov:2011rq,ATLAS:2012an} have been used for the charge asymmetry,
\begin{equation}\label{asym}
A_{\ttt}=\frac{\sigma_{+}-\sigma_{-}}{\sigma_{+}+\sigma_{-}}~.
\end{equation}
At the Tevatron, due to the CP invariant initial state, the charge asymmetry corresponds to a forward-backward asymmetry. For the Tevatron, Standard Model predictions are lower than the measured values, while for the LHC they are larger. The only case, in which the deviation from the theoretical prediction is larger than 3$\sigma$, is the CDF analysis \cite{Aaltonen:2011kc} where a cut on $\ttt$ invariant mass $(M_{\ttt})$ smaller than 450 GeV is applied. Recently, new analyses on a larger data set have 
been performed by CDF \cite{Aaltonen:2012it}. The deviation is lower than 3$\sigma$ also with the aforementioned cut. Still, a systematic deviation in the dependence of the asymmetry on $M_{\ttt}$ remains. \\ \\
LO top quark pair production from QCD does not produce asymmetric terms; they arise in QCD at NLO \cite{Kuhn:1998kw}. The denominator of \eqref{asym} (the total cross section) starts at $\mathcal{O}(\alpha_{s}^{2})$, the numerator only at $\mathcal{O}(\alpha_{s}^{3})$. Thus, the leading contribution to the asymmetry is $\mathcal{O}(\alpha_{s})$. Also electroweak corrections contribute to both the numerator and the denominator. If we perform a pure expansion in powers of $\alpha_{s}$ and $\alpha$  of \eqref{asym}, the leading orders for which a complete calculation is available\footnote{At the moment only the contribution to $\mathcal{O}(\alpha_{s}^{2})$ from NLO QCD corrections to the denominator is available, the contribution from NNLO QCD corrections to the numerator is still missing. See \cite{Hollik:2011ps} for details on the expansion.} are: the $\mathcal{O}(\alpha_{s})$ mentioned before, the $\mathcal{O}(\alpha)$ and the $\mathcal{O}(\frac{\alpha^{2}}{\alpha_{s}^{2}})$. They respectively correspond to the asymmetric part (numerator) of $\mathcal{O}(\alpha_{s}^{3})$, $\mathcal{O}(\alpha_{s}^2\alpha)$ and $\mathcal{O}(\alpha^{2})$ divided by the LO cross section.

All these three contributions arise from the $q\bar{q}$ initial state. At any order, the $gg$ initial state does not contribute; if we assume that non-valence quark and antiquark PDFs are equal \cite{Pagani:2012kj}, also the $c\bar{c}$, $s\bar{s}$ and $b\bar{b}$ do not contribute. 
\\
\\
At  $\mathcal{O}(\alpha_{s}^{3})$ asymmetric terms come from the  boxes in the virtual corrections to $q\bar{q}\rightarrow \ttt$ and from the interference between the initial and final state radiation of the $q\bar{q}\rightarrow \ttt g$ process.

In the calculation of $\mathcal{O}(\alpha_{s}^2\alpha)$ asymmetric terms, again, it is useful to separate the QED contribution from the weak contribution. 
The QED contribution, as anticipated in the discussion on the cross section, comes from the boxes in the loop corrections and the interference of initial and final state radiation of gluons or photons.
The squared matrix elements contributing to this order can be obtained just replacing, in all possible ways, one gluon with one photon in the squared matrix elements contributing to $\mathcal{O}(\alpha_{s}^{3})$ (see \cite{Hollik:2011ps}). The only differences in the calculation of $\mathcal{O}(\alpha_{s}^{3})$ and QED $\mathcal{O}(\alpha_{s}^2\alpha)$ of the numerator of \eqref{asym} are the couplings and the color structure. 
Thus, the QED $\mathcal{O}(\alpha_{s}^2\alpha)$ contribution from any $q\bar{q}$ initial state can be obtained multiplying the analogue $\mathcal{O}(\alpha_{s}^3)$ result by the $R_{QED}$ factor,
\begin{equation}\label{factor}
R_{QED}(Q_{q})=\frac{36}{5}Q_{q}Q_{t}\frac{\alpha}{\alphas}~\Longrightarrow\quad R_{QED}(Q_{u})\sim0.2~, \quad R_{QED}(Q_{d})\sim-0.1~.
\end{equation}
This factor is independent of the specific definition of the asymmetry and of the kinematical cuts applied.

The remaining weak contribution originates from the same diagrams, from loop corrections and the real gluon radiation, of the QED contribution, where the virtual photon is replaced by a Z boson.
In this case, the mass of the Z boson does not allow to write this contribution as the $\mathcal{O}(\alpha_{s}^3)$ result multiplied by a simple factor in an exact way.
Especially for the dominant $u\bar{u}$ initial state, weak corrections are smaller than in the QED case.

The asymmetric term of $\mathcal{O}(\alpha^{2})$ comes from the squared amplitude of $q\bar{q}\rightarrow Z \rightarrow\ttt$ and its interference with  the $q\bar{q}\rightarrow \gamma \rightarrow\ttt$ amplitude. Indeed, different couplings for different chiralities induce asymmetric terms also for s-channel diagrams. These corrections are numerically of the same order as the
weak $\mathcal{O}(\alpha_{s}^2\alpha)$ corrections.
\\
\\
In total, the ratio $R_{EW}$ of the electroweak corrections of $\mathcal{O}(\alpha)+\mathcal{O}(\frac{\alpha^{2}}{\alpha^{2}_{s}})$ to the asymmetry and the leading contribution $\mathcal{O}(\alpha_{s})$ is around $20\%$ at the Tevatron. Thus, electroweak corrections reduce the gap between QCD prediction and experimental measurements. The precise value of $R_{EW}$ depends on the renormalization and factorization scale $\mu$, the eventual cuts applied and, very slightly, on the specific definition of the asymmetry.
The largest contribution to $R_{EW}$ comes from QED $\mathcal{O}(\alpha_{s}^2\alpha)$ corrections to the numerator. The contribution from $d\bar{d}$ is, according to \eqref{factor}, negative and partially cancels the positive contribution from $u\bar{u}$.
The cancellation happens also for weak $\mathcal{O}(\alpha_{s}^2\alpha)$ corrections to the numerator.

At the LHC, top quark pair production is dominated by $gg\rightarrow \ttt$, that contributes only to the denominator of \eqref{asym} diluting the contribution of $q\bar{q}\rightarrow \ttt$ in the numerator.  
Increasing the total energy in the center of mass, the ratio of $u\bar{u}$ and $d\bar{d}$ luminosities decreases. Thus, at LHC and especially at 14 TeV the cancellation between electroweak contributions from $u\bar{u}$ and $d\bar{d}$ initial states is larger. Conversely, the cancellation decreases applying cuts on the $\ttt$ invariant mass.
Moreover at higher energies also the electroweak and QCD contribution to the asymmetry from $qg\rightarrow\ttt q$ process is non negligible.
The value of $R_{EW}$, for the LHC, ranges between $15\%$ and $20\%$, depending on the energy in the center of mass and the cuts applied (see \cite{Bernreuther:2012sx}).
As illustrative comparison, the value of $R_{EW}$ without cuts and $\mu=m_{t}$ is equal to 22\% at the Tevatron and 15\% at the LHC 14 TeV.
\section{Conclusion}
Electroweak corrections to the integrated cross section of top quark pair production ($\sim$ -1\% of the LO prediction) are smaller than QCD uncertainties from scale variation and PDFs.
The size of the relative corrections increases in differential distributions involving a very hard top. Probably, with high luminosity it will be possible to test them with experimental data.
The contribution from QCD to the charge asymmetry arises only within a NLO calculation, so the phenomenological impact of electroweak corrections is much larger than in the case of the cross section.
Electroweak contributions increase the QCD prediction for the asymmetry by a factor $\sim$ 1.2 at the Tevatron and by a factor between 1.15 and 1.20 for the different energies of the LHC.

\def\Discussion{
\setlength{\parskip}{0.3cm}\setlength{\parindent}{0.0cm}
     \bigskip\bigskip      {\Large {\bf Discussion}} \bigskip}
\def\speaker#1{{\bf #1:}\ }
\def\endDiscussion{}

\end{document}